# THE POLITICS OF POSTMORTEM PRIVACY

*Mauricio Figueroa*

## *Abstract*

*The persistence of the dead through data has become a defining feature of contemporary informational life. In response, postmortem privacy has emerged as a recognised concept across legal, ethical, and policy debates, challenging the long-standing assumption that privacy protects only the living. Yet while the existence of postmortem privacy is increasingly acknowledged, far less attention has been paid to its internal instability: its scope, justificatory foundations, and uneven articulation across jurisdictions. This piece unearths and distils the internal diversity of the concept by illuminating specific points of tension and conflict that the notion evokes, which are collectively refer to as the 'politics' of postmortem privacy. To do so, this chapter organises existing contributions of legal scholarship, placing them in dialogue with broader cultural, historical and political observations to illustrate the politics of postmortem privacy through three different loci of analysis: the transatlantic divide between European and American approaches, intra-European tensions within data protection governance, and postcolonial and post-authoritarian contexts in the Global South. While existing literature has glimpsed toward the former two, this piece contends that the latter deserves greater attention and inclusion in the debates around privacy and the dead. The piece argues and advances, in continuity with existing scholarship, how postmortem privacy is assembled differently across settings. For postmortem privacy, heterogeneity is not necessarily a doctrinal failure, but a productive register through which societies negotiate memory, dignity, and the governance of data of the dead.*

# THE POLITICS OF POSTMORTEM PRIVACY

*Mauricio Figueroa*[1]

## Introduction

The persistence of the dead through data has become a near-universal condition of contemporary life. Digital data does not fade at death; it endures across platforms, archives, and databases, posing a shared normative and regulatory problem across cultures, regions, and jurisdictions. The growing consolidation of postmortem privacy in academic debates, legislation, ethics, and policy discussions speaks precisely to the impossibility of ignoring this phenomenon. Postmortem privacy has played an important role in unsettling the *demodé* assumption that privacy belongs only to the living. Yet while this corrective move has been decisive, far less attention has been paid to the internal instability of postmortem privacy itself: what it entails, when it applies, how it operates, and for whom it is ultimately invoked.

Building on existing scholarly contributions, this piece interrogates how postmortem privacy emerges from the interplay and collision of legal traditions, institutional arrangements, political economies, and culturally embedded understandings of death, memory, and justice.

To substantiate this claim, this piece identifies and analyses three loci of analysis through which postmortem privacy is currently being constituted and disputed. The first concerns the transatlantic divide between European and American approaches. While both address privacy after death, they do so through markedly different legal imaginaries: European frameworks tend to situate postmortem privacy within dignity-oriented and testamentary traditions, whereas the American model privileges fiduciary logics, contractual ordering, and platform-driven governance. The second locus of conflict unfolds within Europe itself. Here, the strategic silences of EU data protection law and the deference to Member State regulation expose unresolved tensions between public regulatory authority and private ordering (particularly at the intersection of data protection and family law) raising questions about where, and whether, the dead can be located within the supranational European legal order. The third locus shifts attention beyond the Global North to postcolonial and post-authoritarian contexts in the

[1] Dr Mauricio Figueroa, Durham University, Law School. mauricio.figueroa@durham.ac.uk. Thanks to Tal Morse, Edina Harbinja, Patrick Stokes and Andelka Phillips for their very useful comments.

Global South, where postmortem privacy intersects with projects of memory, truth, and accountability, and where demands for visibility may conflict with presumptions of withdrawal or erasure. Prior scholarship has observed the first two loci, though not through the analytical framework advanced here. Yet, the third locus, pertaining to Global South and postcolonial tensions, has been largely overlooked. Notably, all three points of analysis reveal different facets of a single phenomenon.

Before moving forward, it is useful to clarify what this piece means by *politics*. The term should not be read as a claim that this analysis belongs to any particular school of legal thought. Notably, legal scholars have long observed that law does not operate in isolation from the social worlds in which it is embedded. Legal realism, in clear clash against the received wisdom of legal positivism, emphasised that legal rules and decisions are shaped by social context.[2] Importantly, today it is commonplace in legal education and scholarship to discuss policy considerations as part of legal reasoning.[3] Other approaches have gone further, arguing that law frequently advances particular social or political objectives,[4] and that the line between law and politics is often difficult to maintain. The invocation of *politics* in this piece is not meant to situate the analysis particularly within any one of these traditions, though it is broadly consonant with their insights. Instead, I use the term to capture the structured tensions, competing interests, and normative orientations that emerge within and through a concept. If politics describes the processes through which competing claims about what (or whose) interests should prevail and how these are negotiated, then the domain of postmortem privacy contains its own internal politics.

The contribution proceeds as follows. It begins by introducing the notion of postmortem privacy and explaining why its very emergence constitutes a challenge to the traditional view of privacy as a concern of the living alone. It then examines the limits of uniformity. The analysis then unfolds the three loci of analysis outlined above. The final section reflects on heterogeneity itself, arguing that diversity in postmortem privacy regimes should not be read as a failure of law, but as a revealing feature of a domain in which questions of dignity, memory, power, and governance remain fundamentally unsettled.

---

[2] See, generally, Joseph William Singer, 'Legal Realism Now' (1988) 76 Calif. L. Rev. 465.

[3] Amy Kapczynski, 'Realism, Law and Economics, and LPE Now' (2026) 93 U. CHI. L. REV. 439, 445.

[4] See, generally, Roberto Mangabeira Unger, 'The Critical Legal Studies Movement' [1983] Harvard law review 561; Mark Tushnet, 'Critical Legal Studies: A Political History' (1990) 100 Yale Lj 1515 (explaining how and why critical legal studies scholarship is less an intellectual movement than it is a way to articulate and contain a set of specific political views within American legal academia).

## I. The emergence of postmortem privacy

Do the dead retain a claim to privacy? For much of legal history, and across different jurisdictions, the prevailing response had been mostly in the negative.[5] The reconfiguration of human lives through the use and adoption of digital technologies and the proliferation of digital data, characterised by its resistance to decay and its persistent replicability, challenges conventional understandings. The digital data that outlives individuals is closely linked and, in many cases constitutes, the very essence of our personal lives. In this section, this contribution traces the conceptual emergence of postmortem privacy within legal and scholarly discourse, attending to the key interventions that have shaped the field. It then considers how a range of legal systems have incorporated the principle, before turning to the contestations and debates that have emerged in the process.

### *A. Theoretical foundations*

Postmortem privacy exists today as a theoretical and legal construct with increasing normative traction.[6] Yet its theoretical acceptance has not been automatic; rather, it has emerged through contestation against inherited assumptions about the boundaries of legal subjectivity. Sources have not been able to identify the actual emergence of *actio personalis moritur cum persona* (a personal right of action dies with the person). It is certainly not a roman law principle despite being a Latin construct, and it is not directly stated as such in classical Roman legal sources.[7] It appears instead as a creature of the common law, with sources variously locating its emergence in the late fifteenth or early seventeenth century.[8] What is relatively clearer is the principle's role in differentiating personal from proprietary actions; wherein only the latter is deemed transmissible beyond death.[9] This means that causes of action connected to tangible property or claims of monetary value generally survive the death of a party and are transmissible to their estate,[10] whereas causes of action that are purely personal in nature extinguish upon death.

[5] See, in US law, Atkinson v John E Doherty & Co 121 Mich 372 (1899); in the European context, The Estate of Kresten Filtenborg Mortensen v Denmark (dec) App no 60599/00 (ECtHR, 15 May 2006). In the latter, the Court emphasized that: 'it would stretch the reasoning developed in this case-law too far to hold... that DNA testing on a corpse constituted interference with the Article 8 rights of the deceased's estate. Accordingly, the Court considers that there has been no interference with the rights of KFM's estate for the purposes of Article 8 § 1 of the Convention.'

[6] See, generally, Edina Harbinja, *Digital Death, Digital Assets and Post-Mortem Privacy: Theory, Technology and the Law* (Edinburgh University Press 2022) (providing a robust explanation and justification of postmortem privacy in line with claims of autonomy, dignity and technological development). Elaine Kasket, *All the Ghosts in the Machine: The Digital Afterlife of Your Personal Data* (Hachette UK 2019) (providing a technocultural account of the axis privacy, death and continuing bonds).

[7] For instance, this principle does not appear in the Corpus Iuris Civilis. See T Mommsen and others (eds), Corpus Iuris Civilis (Weidmann).

Such categorical distinctions were perhaps workable in an era when the self was presumed to die with the body. But the contemporary informational subject does not expire so neatly. The rise of digital platforms and storage infrastructures, hallmarks of the information economy, has reconfigured the (in)materiality of human remains. No longer reducible to corporeal fatality, the self persists in code, in networks, in data points scattered across platforms whose logics permeate our societies and exceed individual control.

For readers in law and technology or cultural studies who are approaching this topic for the first time (while those already working in the field will be familiar with it), the work and theoretical contributions of Edina Harbinja offer a rich and solid point of entry into the study of postmortem privacy. In 2013, in collaboration with Lilian Edwards, she introduced the term *post-mortem privacy* (hyphenated),[11] denoting the right of a person to preserve and control what becomes of his or her information, dignity,

---

[8] AWB Simpson, 'Transmission of Liability on Death' in AWB Simpson (ed), *A History of the Common Law of Contract: The Rise of the Action of Assumpsit* (Oxford University Press 1987) <https://doi.org/10.1093/acprof:oso/9780198255734.003.0015> accessed 1 March 2026 (analysing how the maxim was quote in 1486, before the common assumption of 1612).

[9] Id

[10] Cf, Daniel Seng and Kelvin FK Low, 'Data Objects: New Things or No-Thing More than Ignis Fatuus?' (2025) 17 Law, Innovation and Technology 1 (Noting that an asset need not possess monetary value to qualify as such. The authors argue that equating assets with value is a misconception, as there are numerous examples of assets with no value, or even negative value, that nonetheless fall within the ambit of property rights. By analogy, photographs of a holiday may constitute intellectual property rights despite lacking any monetary value). Seng and Low's argument is, to a great extent, a response and criticism to the Law Commission's' report on Digital Assets and the derived Digital Assets Act. See, by contrast, Law Commission, Digital Assets (Law Com No 411, 2023); Property (Digital Assets) Act 2024.

[11] In this work, I deliberately adopt the unhyphenated form: postmortem privacy. This is a symbolic choice, reflecting a broader cultural shift toward normalisation and conceptual maturity, echoing the lexical evolution and stabilisation of terms such as multicultural, postgraduate, interdisciplinary, as exemplified now-familiar terms email, database, or chatbot. The dropping of the hyphen may signal both normalisation and an epistemic consolidation of the term within legal and critical discourse.

integrity, secrets or memory after death.[12]

Harbinja's subsequent work grounds postmortem privacy in the foundational principles of autonomy, dignity, and freedom, reframing it not as an imposed and artificial novel right but as a necessary extension of existing normative commitments.[13] Connected and notable efforts by Michael Birnhack and Tal Morse have provided an influential mapping for engaging with *digital remains*, offering a taxonomy that distinguishes between intangible items, property-related information, intellectual property, and personal data.[14] Their intervention underscores a critical point: although all these data types are linked to the deceased, they are governed by disjointed legal regimes: inheritance law, intellectual property law, and data protection law, respectively, none of which fully captures the continuity of the digital self they implicate, but depend on one another to sustain a legal governance framework.

The conceptual foundation of postmortem privacy has found resonance in further theoretical elaborations. For instance, J. C. Buitelaar interrogates the extension of informational self-determination into posthumous contexts, asking whether this principle may have validity with regards to the digital persona that survives physical death.[15] In the United States, Anita Allen and Jennifer Rothman have advanced detailed accounts of how postmortem privacy is already latent within various legal domains (administrative law, medical confidentiality, the right of publicity, among others) and how these domains may serve as sites for its future development.[16] Their work pushes back against the received wisdom that American jurisprudence offers no privacy protection to the dead.

Additionally and importantly, scholars have consistently mapped quantitative aspects of postmortem privacy that add texture to the theoretical picture. For instance, Tal Morse, in coauthorship with Michael Birnhack in the case of Israel,[17] and later with Edina Harbinja and Lilian Edwards in the UK,[18] has revealed through surveys in these two countries a marked divergence between users' normative commitments and their practical behaviours: while users overwhelmingly express a desire to control their digital legacies, few engage with the tools designed to facilitate such control. More paradoxically still, some may wish to enable posthumous access to their data but, through inaction, inadvertently foreclose it. This inverted paradox speaks precisely to the difficulty of designing and adopting tools around death within systems optimised for the living.

---

[12] Lilian Edwards and Edina Harbinja, 'Protecting Post-Mortem Privacy: Reconsidering the Privacy Interests of the Deceased in a Digital World' (2013) 32 Cardozo Arts & Ent. LJ 83.

[13] See, generally, Edina Harbinja, 'Post-Mortem Privacy 2.0: Theory, Law, and Technology' (2017) 31 International Review of Law, Computers & Technology 26.

[14] Michael Birnhack and Tal Morse, 'Digital Remains: Property or Privacy?' (2022) 30 International Journal of Law and Information Technology 280 (establishing four different categories of digital remains, differentiating between those elements of an economic value and privacy-related remains).

[15] See, generally, JC Buitelaar, 'Post-Mortem Privacy and Informational Self-Determination' (2017) 19 Ethics and Information Technology 129.

[16] Anita L Allen and Jennifer E Rothman, 'Postmortem Privacy' (2024) 123 Michigan Law Review 285.

[17] Tal Morse and Michael Birnhack, 'Digital Remains: The Users' Perspectives' *Digital*

As the existing body of scholarship makes it increasingly clear, postmortem privacy now stands as a recognised conceptual and normative category. Yet its precise boundaries, both in terms of substantive content and normative extension remain undefined. Although there is a shared semantic understanding that privacy interests persist after death, the *whens*, the *whos*, and the *hows* of postmortem privacy remain far from unified. This indeterminacy should not be read as a conceptual flaw; rather, it signals the theoretical richness and structural complexity of the terrain. Furthermore, the ambiguity reflects the vitality of an emergent field whose contours are still in the process of consolidation, shaped through contestation across legal doctrine and scholarship, cultural practices, and technological affordances. In the sections that follow, I bring these perspectives into dialogue to render visible the politics that animate postmortem privacy, illustrating how these nascent frameworks give rise to new tensions: between state imperatives and corporate interests, across cultural contexts, and among competing claims to collective memory.

### *B. Articulation across different jurisdictions*

Privacy and data protection are not synonyms, despite being related and intertwined. Mireille Hildebrandt offers a useful lens through which to differentiate them, at least provisionally: privacy is an opacity right, concerned with preventing others from knowing about oneself across broader realms of human life, whereas data protection is more closely aligned with a transparency right, whereby the individual, vis-à-vis an organisation, seeks to know what is being done with their data.[19] Postmortem privacy debates have largely, albeit not exclusively, been articulated within this latter terrain. The

---

*afterlife* (Chapman and Hall/CRC 2020).

[18] Edina Harbinja, Tal Morse and Lilian Edwards, 'Digital Remains and Post-Mortem Privacy in the UK: What Do Users Want?' [2025] International Review of Law, Computers & Technology 1.

[19] See, generally, Mireille Hildebrandt, *Law for Computer Scientists and Other Folk* (Oxford University Press 2020).

question of whether the privacy of the dead deserves legal protection has assumed renewed relevance and depth with the advent of the General Data Protection Regulation (GDPR).[20]

Recital 27 of the GDPR makes clear that the Regulation does not extend to the personal data of the deceased but leaves Member States free to articulate the matter within their own domestic legal orders.[21] As David Erdos discusses, the European landscape is marked by terminological and conceptual diversity that long predates the GDPR. The varied deployment of categories such as *individual*, *natural person*, or *person in physical terms* complicates the project of locating the deceased within a coherent framework of data protection.[22]

Erdos traces the contours of a post-GDPR paradigm against postmortem privacy, highlighting that inconsistency is the norm. Some jurisdictions afford a measure of protection, but no jurisdiction has gone so far as to extend the law's reach fully and indefinitely beyond death.[23] Erdos demonstrates that protection is typically circumscribed by temporal limits. In some instances, these limits are explicit, eg. 10 years, as in Estonia; in others, they operate implicitly, limited to the lifespan of particular actors such as heirs, as in Italy, Portugal, and Spain. This implies that once the relevant actor dies, the architecture of postmortem privacy dissolves. Even those states that impose broader obligations (such as Denmark and Iceland) tend to define the duration of such duties in specific terms.[24]

Outside Europe, different logics prevail. In the United States, the Uniform Law Comission (ULC) has introduced the Revised Uniform Fiduciary Access to Digital Assets Act (RUFADAA), which gives Internet users the power to plan for the management and disposition of their digital data.[25] It endows fiduciaries with authority to manage digital assets and electronic communications as they would tangible property or financial accounts, while simultaneously empowering custodians to mediate such

---

[20] Regulation (EU) 2016/679 of the European Parliament and of the Council of 27 April 2016 on the protection of natural persons with regard to the processing of personal data and on the free movement of such data, and repealing Directive 95/46/EC (General Data Protection Regulation) [2016] OJ L119/1 (GDPR).

[21] GDPR, Recital 27

[22] David Erdos, 'Dead Ringers? Legal Persons and the Deceased in European Data Protection Law' (2021) 40 Computer Law & Security Review 105495.

[23] *ibid.*

[24] *ibid.*

[25] See, generally, Revised Uniform Fiduciary Access to Digital Assets Act 2015 (US). See also Lucien Castex, Edina Harbinja and Julien Rossi, 'Défendre Les Vivants Ou Les Morts?: Controverses Sous-Jacentes Au Droit Des Données Post Mortem à Travers Une Perspective Comparée Franco-Américaine' (2018) 210 Réseaux 117 (comparing the US and France approaches and their contrasting logics).

access in line with the deceased user's reasonable expectations of privacy.[26]

The largest common law jurisdiction, India, conflates postmortem and incapacity protections. Its regime permits the *data principal* (the analogue of the data subject) to designate ex ante another individual who will exercise data protection rights upon the principal's death or incapacity, thereby introducing a so-called *right to nominate* into the data protection landscape.[27]

Looking to other regions of the Global South, distinct configurations of postmortem privacy emerge. Argentina, for instance, admits only a narrow accommodation: heirs may exercise rights of access to data of the deceased, but there is no equivalent recognition of rights such as rectification.[28]

China, in turn, provides varying degrees of post-mortem privacy protection. Article 49 of the Personal Information Protection Law provides that, upon the death of a natural person, their close relatives may, for their own lawful and legitimate interests, exercise the rights to consult, copy, correct, and delete the deceased's relevant personal information (unless the deceased arranged otherwise prior to death).[29] Less so, however, is the question of whether protections relating to likeness and personal image under the Civil Code of the People's Republic of China extend beyond death,[30] an ambiguity that has been noted by commentators in Chinese legal scholarship.[31]

---

[26] Ibid. s 4.

[27] Digital Personal Data Protection Act 2023 (India) s 4.

[28] Ley 25.326 de Protección de los Datos Personales [Personal Data Protection Law] (Argentina) art 14.

[29] Personal Information Protection Law 2021 (China) art 49, English Version translated by https://personalinformationprotectionlaw.com/ accessed 8 March 2026 (noting that while this translation is used for reference, the original Mandarin text remains the sole authoritative version).

[30] Civil Code of the People's Republic of China 2020 (China) art 1018. Unless otherwise stated, references to the Civil Code are to the translation provided by the World Intellectual Property Organization (WIPO) https://www.wipo.int/wipolex/en/legislation/details/21757 accessed 8 March 2026. (As noted above, the original Mandarin text remains the sole authoritative version).

[31] Kwan Yiu Cheng, 'The Law of Digital Afterlife: The Chinese Experience of AI Resurrection and Grief Tech' (2025) 33 International Journal of Law and Information Technology (providing a discussion of Chinese law in relation to generative technologies and digital resurrection projects. Surprisingly, however, the article does not engage with the earlier body of scholarship that identified and analysed the digital afterlife industry and articulated frameworks for postmortem privacy, even though several elements of its narrative and analytical lens parallel that prior literature. See, eg, Carl Öhman and Luciano Floridi, 'The Political Economy of Death in the Age of Information: A Critical Approach to the Digital Afterlife Industry' [2017] 27 Minds and Machines 639 [coining the term, and explaining the emergence of, the Digital Afterlife Industry]; See further Edina Harbinja, Lilian Edwards and Marisa McVey, 'Governing Ghostbots' [2023] 48 Computer Law & Security Review 105791 <https://doi.org/10.1016/j.clsr.2023.105791>; See also Mauricio

The catalogue of relevant provisions continues to grow, and further legislative interventions may yet be forthcoming. What is increasingly urgent, however, is not a more exhaustive exercise in comparative legislation, but an inquiry into the assumptions and understandings that animate these divergent approaches. The next section examines three loci of tension through which these dynamics become particularly visible.

## II. Postmortem privacy: a defying principle.

Despite its growing legal recognition and increasing scholarly discussion, postmortem privacy evokes contestation at different levels. Externally, postmortem privacy emerged not only as a new object of regulation, but as a form of contestation in its own right, one that unsettles long-standing assumptions within legal tradition, where the dead were conventionally, and erroneously, understood as beyond protection.[32]

Postmortem privacy therefore evokes contestation on two fronts. First, as a conceptual construct, postmortem privacy operates as a critical device for contesting and debunking *demodé* legal assumptions about the status and significance of the dead. In this respect, postmortem privacy functions externally as a challenge to dominant narratives within legal discourse, challenging the notion that the dead cease to have any sort of protection as to their privacy. Secondly, postmortem privacy is also internally fractured. Its emergence does not signal the consolidation of a coherent vision of protection for the dead, nor does it reflect a settled institutional or societal consensus as to its meaning or scope. Instead, it is characterised by competing claims and divergent normative commitments that render its contours diffuse and unstable. While the first axis of contestation (the external challenge to inherited legal assumptions) has been widely explored in legal scholarship, the second axis, concerning internal heterogeneity, remains comparatively underexamined. It is within this space of internal divergence that this contribution intervenes. The ensuing sections aim to unpack the internal diversity of postmortem privacy to explain how from the inside it holds

---

Figueroa-Torres, ‘Affection as a Service: Ghostbots and the Changing Nature of Mourning’ [2024] 52 Computer Law & Security Review 105943 <https://doi.org/10.1016/j.clsr.2024.105943>; See further Carl Öhman, ‘The Afterlife of Data: What Happens to Your Information When You Die and Why You Should Care’ *The Afterlife of Data* [University of Chicago Press 2024]; See further Tomasz Hollanek and Katarzyna Nowaczyk-Basińska, ‘Griefbots, Deadbots, Postmortem Avatars: On Responsible Applications of Generative AI in the Digital Afterlife Industry’ [2024] 37 Philosophy & Technology 63 <https://doi.org/10.1007/s13347-024-00744-w>.

[32] See, above, Part. I. A.

different expressions and visions.

### *A. Situated legalities*

Postmortem privacy, as illustrated in the previous section, has been incorporated in different legal systems with different provisions. But legislative convergence does not necessarily entail normative uniformity. Divergent normative justifications are not a threat to a principle's legitimacy; it is a symbol of its richness and how cultural and historical contexts intersect with the law.

While Berkowitz et al argue that legal transplants are common but often produce weaker legal institutions when they are adopted without local demand or familiarity,[33] comparative legal theorist Pierre Legrand takes a more radical position. Legrand insists that a legal rule cannot be meaningfully abstracted and transplanted from one cultural lifeworld to another becomes particularly relevant.[34] For Legrand, the rule is never simply a neutral, portable artefact; it is a situated expression, a condensation of the symbolic and institutional grammar of a legal culture. It is against this backdrop that, despite sharing a common conceptual vocabulary, postmortem privacy cannot really operate as a stable or universally intelligible category across legal systems. Instead, it emerges as a site where cultural values, economic interests, and institutional frameworks interact to produce distinct conditions of meaning and application.

The multiplicity of meanings and rationales is not something exclusive of postmortem privacy. In fact, privacy at large has a similar configuration. Daniel Solove, drawing on Wittgenstein's notion of "family resemblances," contends that privacy is not best understood as a singular, unified concept but as a cluster of related ideas that share overlapping features.[35] Solove's family resemblance approach encourages scholars to focus on how privacy is invoked, protected, and violated in practice. This non-essentialist account mirrors Legrand's insistence that legal concepts cannot be abstracted from the cultural and institutional contexts in which they acquire meaning. Both argue, in their own way, that meaning emerges from use, interpretation, and institutional embedding rather than from formal abstraction. If privacy itself has no singular essence, then postmortem privacy (an even more context-sensitive and contested subdomain) cannot plausibly be expected to travel unchanged across legal systems.

---

[33] See, generally, Daniel Berkowitz, Katharina Pistor and Jean-Francois Richard, 'The Transplant Effect' (2003) 51 Am. J. Comp. L. 163.

[34] Pierre Legrand, 'The Impossibility of "Legal Transplants"' (1997) 4 Maastricht journal of European and comparative law 111 (critiquing the adoption of foreign legal principles disregarding local legal culture and dynamics).

[35] See, generally, Daniel J Solove, 'Conceptualizing Privacy' (2002) 90 Calif. L. Rev. 1087.

### *B. Lack of stability of the principle*

The conceptual terrain of postmortem privacy is marked by tensions that arise from the multiplicity of stakeholders involved, but also from the entanglement of legal norms with cultural practices, institutional histories, and trajectories of political economy. It is this very entanglement that gives the concept both its traction and its instability across legal systems and normative orders.

Importantly, my argument should not be conflated with an account of legal pluralism. While postmortem privacy may indeed be shaped by overlapping normative domains (including customary, religious, Indigenous, and state-based legal orders) my concern is not with the multiplicity of legal systems as such, but with the points of tension that unfolds within positive law. Even where postmortem privacy is codified in legislation, apparent convergence at the textual level often masks deeper divergences in justificatory frameworks, institutional rationalities, and philosophical commitments.

If privacy as a whole is a concept marked by shifting boundaries and plural meanings,[36] then postmortem privacy is especially thought-provoking for precisely that reason. The internal diversity of postmortem privacy stands in marked contrast to the relatively settled nature of other legal constructions. Consider, for instance, the notion of *trust* in corporate law, which rests on well-defined equitable principles; or *patent* in intellectual property law, which is underpinned by a broadly accepted instrumental rationale centred on innovation incentives. Similarly, *income tax*, while politically contentious, rarely reopens foundational questions regarding its legitimacy as a fiscal mechanism. Even *adoption* in family law, though shaped by contestations around eligibility and access, tends to cohere around a shared understanding of child protection and welfare.

Postmortem privacy, by contrast, remains a site of active and unresolved tension, touching upon complex questions of personhood, memory, relationality, and the persistence of identity beyond death. It does not have the normative closure and the incontestability at the level of foundational principle that may be found in other realms. It is precisely this conceptual volatility that demands a closer examination of the normative logics

[36] See, generally, Julie E Cohen, 'Turning Privacy Inside Out' (2019) 20 Theoretical Inquiries in Law 1 <https://doi.org/10.1515/til-2019-0002>.

underpinning postmortem privacy.

Rather than presuming a singular normative foundation, the following section identifies three distinct loci through which postmortem privacy is constructed and contested.

## III. Three loci of tensions

In this section, this contribution will map three sites of contestation referring to divisions by borders, cultures and economic logics.

### *A. Transatlantic divide*

As Castex et al have observed, the GDPR and RUFADAA both address the question of privacy after death, yet they enact different imaginaries.[37] The GDPR, through its carefully placed silences, invites Member States to craft protections that extend privacy into the posthumous domain. RUFADAA, in contrast, provides a scaffold for American state legislatures to institutionalise fiduciary control over the digital remains. Reading across the two models reveals less a convergence than a bifurcation: each locates authority differently, each privileges distinct institutional interlocutors.

European jurisdictions tend to privilege wills and testamentary directives, embedding postmortem privacy within a dignity-oriented narrative that foregrounds autonomy and freedom of choice.[38] The American approach treats digital remains less as a question of dignity and more as a contract-driven problem of managing externalities and the costs of accessing to digital remains, aligning with the transactional logics of inheritance.[39] The American model pursues efficiency through the terms of service and proprietary tool of platforms, shaping the contours of what counts as permissible.

European jurisdictions that have enacted postmortem privacy legislation place the deceased at the centre of their normative architectures. The language of dignity, autonomy, and freedom pervades these frameworks, even when articulated through the doctrinal idioms of inheritance law and civil codes. The values are not incidental; they structure the very terms of engagement, casting the deceased as a continuing subject of legal concern.

---

[37] See, generally, Castex, Harbinja and Rossi (n 25).

[38] Harbinja (n 13) (providing an early comparison between the European jurisdictions and the American approach).

[39] The American approach tends to conceptualise postmortem digital interests through contractual allocation and cost management rather than through the language of dignity or continuing personality. See Seng and Low (n 10) (noting that the notion of property conceals multiple meanings, including both exclusionary in rem rights over things and obligation-based in personam rights conceived as assets or transferable wealth).

While both the European and American approaches confront the problem of privacy after death, they do so by enacting different legal imaginaries. The GDPR's architectural restraint preserves space for Member States to embed postmortem privacy primarily through the grammar of dignity-oriented and testamentary frameworks. RUFADAA, by contrast, channels posthumous digital governance through fiduciary structures and the contractual architectures of platforms. This contrast underscores that postmortem privacy is not a pre-given legal category, but an outcome of how different legal cultures organise authority, values, and institutional competence in response to the same regulatory problem.

The fact that major platforms such as Google and Facebook formally endorsed RUFADAA through letters of support is itself analytically revealing.[40] That endorsement signals a close alignment between the statute's fiduciary and contract-based architecture and platform interests. By framing postmortem privacy as a matter of fiduciary administration mediated through terms of service, RUFADAA privileges procedural ordering and institutional manageability over open-ended dignity-based claims, situating postmortem privacy within the familiar logics of platform policy. That European approaches to post-mortem privacy speak to different interests should not, in this light, come as a surprise. Once postmortem privacy is understood not as a pre-given legal template but as an outcome of institutional arrangements, regulatory traditions, and political-economic conditions, divergence becomes the expected rather than the anomalous result.

### *B. Intra-European divide*

The European legal order's treatment of postmortem privacy is telling not so much for what it declares as for what it withholds. Its carefully calibrated silences draw attention to an unresolved contestation between public authority and private ordering, revealing the extent to which postmortem privacy remains a negotiated space within the European legal imagination. If, as Legrand contends, European legal systems retain distinct conceptions of public authority, private autonomy, and institutional legitimacy despite formal harmonisation,[41] the GDPR embodies a preserved space for divergent national articulations of postmortem privacy, rather than gesturing towards a uniform European settlement.

---

[40] Google, 'Letter of Support: Google for RUFADDA' (Dated October 13, 2015, Uniform Law Commission Enactment Kit,); Facebook, 'Letter of Support: Facebook' (Dated October 12, 2015, Uniform Law Commission Enactment Kit).

[41] See, generally, Pierre Legrand, 'European Legal Systems Are Not Converging' (1996) 45 International & Comparative Law Quarterly 52.

There is a structural peripherality inscribed in the EU's governance architecture particularly relevant for postmortem privacy. Although data protection lies firmly within the Union's regulatory remit, anchored in Article 8 of the Charter,[42] and given concrete form by the GDPR, postmortem privacy remains conspicuously relegated to Member State discretion. This relegation cannot be explained as a matter of procedural convenience alone. Rather, it points to an epistemological boundary. Ambiguity and subsidiarity operate here as boundary-maintaining devices, preserving constitutional identity by allocating different normative domains to different levels of governance. As one moves deeper into the architecture, data protection assumes a distinctly public character suited to supranational regulation, while questions surrounding death gravitate towards family law and its normative commitments to the private sphere. Whereas the transatlantic divide reflects political economy and corporate interests, the intra-European divide reveals a more fundamental uncertainty about where, institutionally and normatively, the dead ought to be located.

A recurrent claim within EU legal discourse is that the European legal order has transcended the traditional distinction between public and private law, such that the absence of a meaningful divide is treated as an ontological feature of EU governance.[43] Yet this account has become increasingly unstable. An alternative line of argument has sought to 'rediscover' the public/private distinction within European private law, proceeding from the epistemological premise that the divide has never disappeared, but has instead been obscured by dominant modes of integration and harmonisation.[44] Tension over postmortem privacy thus does not merely concern the scope of protection for the dead, but operates as a struggle over where the line between public authority and private ordering ought to be drawn (and who gets to draw it).

More specifically, the deferral of postmortem privacy to national legislatures reflects unspoken hierarchies within the European data protection project itself. What presents as technical discretion also functions as a mechanism for stabilising which values (such as autonomy, efficiency, and harm prevention) are foregrounded, and which (such as dignity, memory, and legacy) are backgrounded or deferred. From this perspective, divergence in postmortem privacy protections across GDPR jurisdictions less an incidental consequence of decentralisation than it is a symptom of more foundational tensions within the European legal order. In particular, it reveals a persistent discomfort with treating death and its digital remains as matters of supranational governance rather than as questions anchored in private and familial spheres.

---

[42] Charter of Fundamental Rights of the European Union [2012] OJ C326/391, art 8.
[43] Martijn W Hesselink, 'Knowing EU Law' (2024) 26 Cambridge Yearbook of European Legal Studies 155, 175 <https://doi.org/10.1017/cel.2025.1>.
[44] *Id*

Historically, death has been understood as a private concern, structured through practices of mourning, inheritance, and remembrance that have largely resisted supra national law intervention. Even in those Member States that recognise forms of postmortem privacy within the GDPR's orbit, such recognition is typically mediated through the doctrinal architectures of civil law and articulated via inheritance law frameworks. For instance, Catalonia's *Llei de voluntats digitals* (Law on Digital Wills) and France's *Loi pour une République numérique* (Law for a Digital Republic) have incorporated and recognised postmortem privacy concerns.[45] Both laws allow individuals to decide during life what happens to their digital data after death, such as social media accounts. Both are inscribed in the register and grammar of inheritance law, not data protection. The deceased is treated primarily as a de cujus (the originator of a succession) rather than as a continuing data subject. To observe this is not to privilege one doctrinal paradigm over another, but to underscore that the tension between data protection and inheritance remains unresolved, and that postmortem privacy continues to mark a fault line in the allocation of normative and institutional authority.

### *C. Global South divide*

Less scholarly attention has been paid to the configuration of postmortem privacy in the Global South. It emerges as a locus of friction wherein competing epistemologies of memory, justice, and historical reckoning come into contact with legal and technological regimes that presuppose a liberal individualist subject. In much of the Global South, postmortem privacy debates are not as developed as in GDPR-derived legislations that account for the dead, or as detailed as in the American RUFADAA. In the Global South, postmortem privacy is actively reconfigured through the legacies of colonialism, dictatorships and structural silencing.

This post-dictatorial and post-authoritarian context in the Global South may influence how those populations make sense of postmortem privacy. Colonial and dictatorial regimes did not simply govern populations; they structured archives, inscribed narratives, and monopolised the machinery of memory.[46] To control memory was to control the terrain on which legitimacy

---

[45] See, generally, Llei 10/2017 de les voluntats digitals [Law 10/2017 on Digital Wills] (Catalonia, Spain); Loi n° 2016-1321 du 7 octobre 2016 pour une République numérique [Law No 2016-1321 of 7 October 2016 for a Digital Republic] (France).

[46] Michel-Rolph Trouillot, *Silencing the Past: Power and the Production of History*

was constructed and contested. In this context, the archival impulse and the need to access data that persist across many postcolonial societies is not really a residual effect of trauma; it is a strategic response to historical erasure.

As Priscilla Hayner has observed, truth commissions typically ground their work in the large-scale collection of testimony from victims, witnesses, and, at times, perpetrators.[47] In post-authoritarian and post-colonial contexts, such practices of documentation and narration do more than generate evidence; they actively constitute collective memory and enable political reckoning. Viewed against this backdrop, postmortem privacy sits in tension with a competing normative imperative: the demand to record, preserve, and render legible the lives and deaths of those who were previously silenced.

In contexts marked by disappearance, censorship, and the systematic denial of truth, the instinct to preserve (even intimate) data acquires political and moral urgency. For instance, Argentina's choice to grant heirs rights of access to the data of the dead, while withholding powers of deletion or rectification,[48] can be read through this register. To delete is to risk repeating the politics of disappearance; to rectify is to risk reinscribing the silences of censorship. Access, by contrast, is a means of countering erasure, of keeping memory open to contestation. Consider the contested afterlives of local political figures or victims of state violence. Deleting or obstructing the access of their data (on social media, in archives, or institutional databases) may be construed not as a matter of ethical stewardship, but as a form of historical sanitisation. Here, the preservation of data becomes a counter-hegemonic act, an effort to hold power accountable by resisting the logics of forgetting. Indeed, memory itself becomes a mode of resistance.

One particularly good example is Article 1392 Bis of Mexico City's Civil Code and its subsequent constitutional outlawing by the Mexican Supreme Court.[49] The provision in the civil code, attempted to incorporate postmortem privacy through the automatic deletion of social media data of the deceased in the absence of instructions to the contrary from the data

---

(Beacon press 2015) 26 (instead of discussing postmortem privacy as such, the author advances that silences enter the process of historical production at four crucial moments: the making of sources, the making of archives, the making of narratives, and the attribution of retrospective significance. In such contexts, postmortem privacy is less a matter of formal legal articulation than of contested memory, archival control, and the afterlives of political violence).

[47] See, generally, Priscilla B Hayner, *Unspeakable Truths : Transitional Justice and the Challenge of Truth Commissions* (2nd ed., Routledge 2010).

[48] See supra Part I. B.

[49] Código Civil para la Ciudad de México [Civil Code for Mexico City] art 1392 bis, last paragaph (noting that this provision was declared unconstitutional by the Mexican Supreme Court, Tesis 1a II/2023 (11a), digital record 2025995, 24 February 2023).

subject when alive.[50] The petitioner, a digital rights organisation, argued that the provision infringes upon the rights to freedom of expression and access to information. While the rationale of the provision seems grounded in the interests of the deceased, the legal tradition of that country suggested alternative grounds for contestation. The Supreme Court declared that provision unconstitutional, finding the rule overly broad and incompatible with constitutional protections.[51] At the heart of the dispute lied a normative conflict between the posthumous privacy and the collective rights to expression, memory, and truth. The Court recognised the potential validity of postmortem data protection but warned that its application must be context-sensitive, particularly where third-party rights or public interest are implicated. A closer look to other experiences in the region seems illustrative to unearth this resistance.

In Argentina, after the return to democracy, the *National Commission on the Disappearance of Persons (CONADEP)* compiled testimonies, documents, and forensic records relating to forced disappearances during the dictatorship. The archives include personal data, testimonies, letters, and evidence from and about the disappeared; most of whom were later declared dead or presumed killed.[52] While these materials involve deeply personal data (sometimes against the wishes of families), they were preserved and published in service of collective memory and justice. Similarly, the *Abuelas de Plaza de Mayo* (Grandmothers of Mayo Square) led efforts to recover the identities of children stolen during the dictatorship. Their work depended on DNA databases, testimonies, and personal records, often of deceased biological parents. The Argentine state authorised the creation of the *National Genetic Data Bank* to support this work,[53] overriding typical privacy concerns in the name of identity and justice. It is not that postmortem privacy is necessarily at odds with this type of initiatives, but the cultural mindset resists an automatic transplantation of postmortem privacy as understood, for instance, in great part of Europe.

Similarly, in Chile, after Pinochet left power, the Chilean government created the *National Commission for Truth and Reconciliation*, tasked with investigating human rights violations that led to death or disappearance. The report compiled testimonies, official records, forensic evidence, and personal histories of the deceased. Families contributed documents, photos, and intimate details. Some families were initially reluctant to share information, fearing reprisal or exposure. Yet the collective goal of truth-telling and historical memory prevailed. The report publicly named victims and offered an official narrative that directly challenged the regime's efforts to suppress or distort memory.[54]

---

[50] *Id*

[51] *Id*

[52] Comisión Nacional sobre la Desaparición de Personas (CONADEP), Nunca Más: Informe de la Comisión Nacional sobre la Desaparición de Personas (EUDEBA 1984) https://www.cultura.gob.ar/media/uploads/lc_nuncamas_digital1.pdf accessed 5 January 2026.

[53] Victor B Penchaszadeh, 'Genetic Testing to Restore the Human Right to Identity in Post-dictatorship Argentina: Ethical, Legal, and Social Issues' *American Journal of Medical Genetics Part C: Seminars in Medical Genetics* (Wiley Online Library 2021).

This striving for memory is especially visible in post-authoritarian and post-conflict settings. In Latin America, families of the disappeared reject the erasure of any digital or material trace, insisting that intimacy must sometimes be sacrificed in pursuit of visibility and justice. In Rwanda and South Africa, truth commissions have relied extensively on postmortem data (testimonies, photographs, forensic records, etc) to reconstruct the conditions of atrocity.[55] The dead, even in their absence, are conscripted into the work of truth-telling. Exposure of the dead, in this sense, is more related to the honour and reintegration of their memory, and less to violations of privacy.

Postmortem privacy is not necessarily against these anxieties, yet the the post-colonial and post-dictatorial mindset prevents the formation of a horizontal principle applicable in the same way as in democracies of the Global North. In the Global South, more often than not, memory is collective, justice is generational, and visibility functions as a safeguard against erasure. Claims over data relating to the dead are therefore less frequently framed in terms of individual agency and control, and more often articulated through vocabularies of truth, accountability, and historical redress.

Against this backdrop, the normative presumption that data ought to recede from public view at death becomes a site of contestation, not because privacy is unvalued, but because its uncritical enactment risks reproducing the very silences that post-authoritarian and post-colonial projects seek to undo.

## IV. Reflections on heterogeneity

Perhaps the universality of postmortem privacy lies not in convergent solutions, but in the shared recognition of a problem that no society can any longer disavow. The persistence of the dead through data has become a near-universal condition of contemporary life, generating a common regulatory and normative challenge across cultures, regions, and jurisdictions.

---

[54] Chilean National Commission on Truth and Reconciliation, Report of the Chilean National Commission on Truth and Reconciliation, vol 1 (University of Notre Dame Press 1993) https://www.usip.org/sites/default/files/resources/collections/truth_commissions/Chile90-Report/Chile90-Report.pdf accessed 6 January 2026.

[55] See, generally, Truth and Reconciliation Commission of South Africa, Final Report (1998); Phil Clark, *The Gacaca Courts, Post-Genocide Justice and Reconciliation in Rwanda: Justice without Lawyers* (Cambridge University Press 2010).

The heterogeneity observed across legal responses cannot be dismissed as a mere by-product of regulatory lag or institutional fragmentation. Instead, it reflects deeper (often tacit) ways in which societies articulate the relationship between the individual, the collective, and the infrastructures of memory through which both are sustained. To speak of postmortem privacy, then, is not simply to invoke a bundle of rights or entitlements, but to engage with the different and sometimes contrasting configurations of they ways the living organise their relationship with the dead.

Importantly, this persistence of the dead through data is neither solely technological nor merely legal. It is irreducibly political. It exposes struggles over authority and power: between the living and the dead, between state institutions and corporate actors, between individual claims and collective projects, between European and American orientations, and between the Global North and the Global South. Across these configurations, the governance of data of the dead reopens unresolved questions of dignity, justice, and memory, questions that may appear tractable in isolation, but become profoundly complex when situated within broader social and institutional contexts.

The absence of a unified framework should therefore not be mistaken for an absence of normativity. On the contrary, it signals the emergence of a plural and contested register: the register of postmortem privacy. Far from being diminished by its diversity, this register is enriched by it, revealing the multiple ways in which societies negotiate care for the dead, the preservation of memory, and the limits of forgetting.

## V. Conclusion

This piece has sought to expand the debates over postmortem privacy, situating it with within wider accounts of cultural and societal tensions. Rather than treating this indeterminacy as a deficiency to be resolved through harmonisation, the chapter has argued that it is precisely through divergence and tension that the significance of postmortem privacy comes into view.

By placing into dialogue a range of otherwise scattered contributions across data protection, comparative law, and memory studies, the chapter has sought to surface the conditions under which postmortem privacy is assembled and contested. Through the identification of three loci of contestation, the analysis has shown that postmortem privacy does not operate as a stable legal recipe capable of uniform application. Instead, it emerges as an outcome of context-specific negotiations shaped by institutional authority, political economy, cultural understandings of death, and competing imperatives of memory and restraint.

The tensions inherent in postmortem privacy are unlikely to be settled through doctrinal adjustments or interpretative efforts alone. The issue occupies a domain that remains conceptually undefined yet broadly recognised. These tensions do not admit of easy resolution, nor should they be expected to.

Sources